\pdfoutput=1

\documentclass[sigconf]{acmart}

\acmConference[ICSE 2024]{46th International Conference on Software Engineering}{April 2024}{Lisbon, Portugal}

\AtBeginDocument{%
  \providecommand\BibTeX{{%
    \normalfont B\kern-0.5em{\scshape i\kern-0.25em b}\kern-0.8em\TeX}}}

\setcopyright{acmlicensed}
\copyrightyear{2024}
\acmYear{2024}
\acmDOI{XXXXXXX.XXXXXXX}

\acmBooktitle{InteNSE '24: ACM 	International Workshop on Interpretability, Robustness, and Benchmarking in Neural Software Engineering,
 April, 2024, Lisbon, Portugal} 
\acmISBN{XXX-X-XXXX-XXXX-X/XX/XX}

\newtheorem{dfn}{Definition}

\begin{document}

\title{Assured LLM-Based Software Engineering}
\author{Nadia Alshahwan}
\orcid{https://orcid.org/0009-0009-4763-0396}
\authornote{Author order is alphabetical. The corresponding author is Mark Harman. }
\author{Mark Harman}
\orcid{https://orcid.org/0000-0002-5864-4488}
\author{Inna Harper}
\orcid{https://orcid.org/0009-0008-9359-0949}
\author{Alexandru Marginean}
\orcid{https://orcid.org/0009-0001-5311-762X}
\author{Shubho Sengupta}
\orcid{https://orcid.org/0009-0007-4204-5185}
\author{Eddy Wang}
\orcid{https://orcid.org/0009-0009-8825-6986}
\affiliation{%
  \institution{Meta Platforms Inc.,}
  \streetaddress{1 Hacker Way}
  \city{Menlo Park}
  \state{California}
  \country{USA}
}
\renewcommand{\shortauthors}{Alshahwan and Harman, et al.}

\begin{abstract}
In this paper we address the following question:
How can we use Large Language Models (LLMs) to improve code independently of a human, while ensuring that the improved code
\begin{enumerate}
\item does not regress the properties of the original code ?
\item improves the original in a verifiable and measurable way ?
\end{enumerate}

To address this question, we advocate Assured  LLM-Based Software Engineering; a generate-and-test approach, inspired by Genetic Improvement. 
Assured LLMSE applies a series of semantic filters that discard code that fails to meet these twin guarantees.
This overcomes the potential problem of LLM’s propensity to hallucinate.
It allows us to generate code using LLMs, 
independently of any human.
The human plays  the role only of final code reviewer, as they would do with code generated by other human engineers.

\mbox{}\\
This paper is an outline of the content of the keynote by Mark Harman at the International Workshop on Interpretability, Robustness, and Benchmarking in Neural Software Engineering, Monday 15th April 2024, Lisbon, Portugal.
\end{abstract}


\keywords{Large Language Models (LLMs), Genetic Improvement (GI), Search Based Software Engineering (SBSE), Llama, CodeLlama, Automated Code Generation}

\maketitle

\newpage
\section{Introduction}

There has been a great deal of recent work on LLM–based code generation \cite{mhetal:LLM-survey,hou:large} and on sub-areas such as Software Testing \cite{wang:LLM-testing-survey}.
We use the term   ``LLM–based Software Engineering'' \cite{mhetal:LLM-survey} to name this rapidly-developing area of Software Engineering.
LLM-based Software Engineering (LLMSE) is any application  in which the software products and/or processes are based on the use of Large Language Models, in the same way that Search Based Software Engineering is any application in which the products and processes are based on computational search \cite{mhamyz:acm-surveys}, 
and the way in which Component-based Software Engineering is based on components \cite{shashank:systematic}. 

We distinguish between online and offline LLMSE.
In online LLMSE, the results of LLM inferences are required in real time. 
For example, for code completion applications, such as CoPilot \cite{copilot2023} and CodeCompose \cite{murali:codecompose}, where the results  are provided directly in the Integrated Development Environment (IDE) editor as the software engineer interacts with it. 
The LLM provides suggested completions of the current code construct, method or fragment being edited in the IDE.

In offline LLMSE, results of inference are not required in real time.
The additional computation time this affords the overall process   allows offline LLMSE to check, and thereby ensure, that the resulting code satisfies verifiable (or testable), measurable assurances, including:

\begin{enumerate}
    \item Computation of metrics to investigate, test and/or verify desirable and automatically-measurable generated code properties. 
    These properties include execution time, code footprint size, dynamic memory consumption, power consumption, scroll performance, and communications bandwidth. 
    These kinds of operational properties are typical of the measurable and improvable characteristics that are targeted by work on Genetic Improvement \cite{Petke:gisurvey}.
    \item Generation  and execution of tests that ensure that the code generated by the LLM does not regress any functional or operational characteristics of the original code base into which it will be deployed.
    \item Post processing to optimize  the code. 
    For example, by running linter processes that catch known issues and fix these. 
    The post processing can include static analyses and manipulations that improve the generated code. 
    \item Re-prompting (for example,  in a chained prompting strategy \cite{wei:chain,yao:react}) to improve the results produced by previous LLM responses. 
    This re-prompting may form part of a Search Based Software Engineering (SBSE) process \cite{mhamyz:acm-surveys} that self-optimizes the prompting strategy.
\end{enumerate}

All of these activities take time, especially those that require rebuilding the system being improved \cite{mh:apr22-keynote}.
However, the additional time available in offline mode affords the ultimate benefit to the assured offline approach: the code presented to the engineer  comes with testable/verifiable measurable guarantees of performance and reliability.
In the online approach, such assurances are limited to those that can be computed in real time, which will not be sufficient for all kinds of assurance that will be useful to an overall  software engineering process.

We discuss applications of this form of LLMSE, and its relationship to Genetic Improvement, concluding with open problems.
We are interested in collaborating with partners in the academic and research communities on these problems.

\section{On/Off Line LLM-based Software Engineering }

This section sets out more formal definitions of LLM-based Software Engineering (LLMSE), 
allowing us to make a more precise distinction between online and offline LLMSE.

\begin{dfn}[LLM-Based Software Engineering (LLMSE)]
{\rm 
LLM-based Software Engineering ({\em LLMSE}) is any application of software engineering in which the products and/or processes are based on the use of Large Language Models.
}
\end{dfn}

\begin{dfn}[LLM Application]
{\rm 
The LLM {\em application} is defined to be any activity that can be improved by the availability of LLM responses.
}
\end{dfn}

\begin{dfn}[LLM Consumer]
{\rm 
The LLM {\em consumer} is defined to be any entity or process that receives the LLM responses.
}
\end{dfn}

Note that the LLM consumer can be a human; a software engineer for example. 
However, the definition does not confine itself to human consumers. 
The LLM consumer can, for example, be another software engineering tool or process.
It could even be another LLM.

The distinction between online and offline LLMSE rests critically on the definition of `real time'.
There are many different ways of defining `real time' \cite{dodhiawala:real,kopetz:real}, 
all of which include some concept of `timeliness', 
but which may include additional attributes such as graceful adaption \cite{dodhiawala:real} that are not relevant here.

We take a utility–led approach to our definition of `real time', that focuses on the consumer of the response.
Essentially, the response is `real time' if its consumer would not behave meaningfully differently, were the response time to be reduced.
This seems to be an elegant way of capturing `timeliness' in a manner suitable for Software Engineering.
More formally:

\begin{dfn}[Real Time]
{\rm 
A result is provided in {\em real time} for a consumer $c$ and application $a$, if and only if $c$ cannot perform $a$ any better when the time spent awaiting the LLM response reduces.
}
\end{dfn}

Notice that the definition of real time is parameterized by the consumer.
This is an important distinction.
A result that is rendered in real time for a software engineer (for example, a code completion result), 
may not be deemed to be real time for a software tool that consumes the response.

The definition of real time is also parameterised by the application to which the LLM response is put.
For example, suppose a software engineer is using the LLM to provide question content for a programming quiz.
The duration deemed acceptable to qualify as `real time' is likely to be considerably longer for this application than the application of code completion in the IDE editor.
Applications that cannot meet the real time constraints for software engineers as consumers need not be performed outside the IDE altogether; the IDE might still prove to be the best deployment route of maximal developer relevance for these applications. 
Nevertheless, such applications  cannot be expected to execute after each keystroke in the IDE's editor.

\begin{dfn}[Online LLM-Based Software Engineering (Online LLMSE)]
{
\label{dfn:online}
\rm 
{\em Online} LLM-based Software Engineering is any application of LLMSE in which the results are always provided to the LLM consumer in real time.
}
\end{dfn}

According to Definition~\ref{dfn:online}, what constitutes online LLMSE depends on the definition of real time. 
This forces us to determine what would constitute a  `reasonable elapsed time' before which the consumer (e.g., a software engineer) becomes `aware' of having to wait.
We leave this definition parameterised in this way so that it can be adapted to different application scenarios, each of which may have  different definitions of real time.
However, for scenarios in which the human is the consumer, 
and the application requires  a response that seems instantaneous to the human, 
we can reasonably  assume that this time limit is bounded above by one second \cite{shneiderman:response}.

Code completion is the archetypal  application of online LLMSE with a human consumer.
Such applications include CoPilot \cite{copilot2023} and CodeCompose \cite{murali:codecompose}.
In these applications, the software engineer begins typing, 
and the LLM is running continuously to recommend completions of the current prefix typed by the human. 
This has to be real time in the sense that the human is unaware of any time spent awaiting the response. 
Any time that the human spends waiting will tend to reduce the performance of the engineer.

Much of the recent research interest and excitement about LLMSE has, hitherto, focused on code completion and most of the impact so far felt by software practitioners has derived from the realm of online LLMSE \cite{mhetal:LLM-survey}.
However, the field is in an embryonic state of development, and the current prevalence of code completion in particular 
(and online LLMSE in general) may shift over time.

In this paper we highlight the need for research on a particular kind of offline LLMSE, {\em assured offline LLMSE}, terms which we now define more formally.

\begin{dfn}[Offline LLMSE]
{\rm 
{\em Offline} LLMSE is defined simply as the opposite of online LLMSE.
That is, offline LLMSE is any application of LLMSE in which the results are {\em not} always provided to the LLM consumer in real time.
}
\end{dfn}

Our distinction between online and offline LLMSE is  similar to the distinction between online and offline learning
\cite{hoi:online}.
As with online learning, it might (at first sight) appear that online LLMSE would {\em always} be preferred over offline LLMSE. 
Indeed, all else being equal, this is definitely the case. 
However, the extra time available to offline LLMSE affords the opportunity to provide assurances, that may be simply impractical to provide in an online setting.
Let us formally define assured LLMSE so that we can clarify the distinction:

\begin{dfn}[Assured LLMSE]
{\rm 
An {\em Assured} ~LLMSE outcome is defined as a (possibly post-processed) LLM response that comes with some verifiable claim to its utility.
In Assured LLMSE, all LLM responses come with such assurances.
}
\end{dfn}

\begin{figure*}[t]
\centerline{\includegraphics[width=0.9\linewidth]{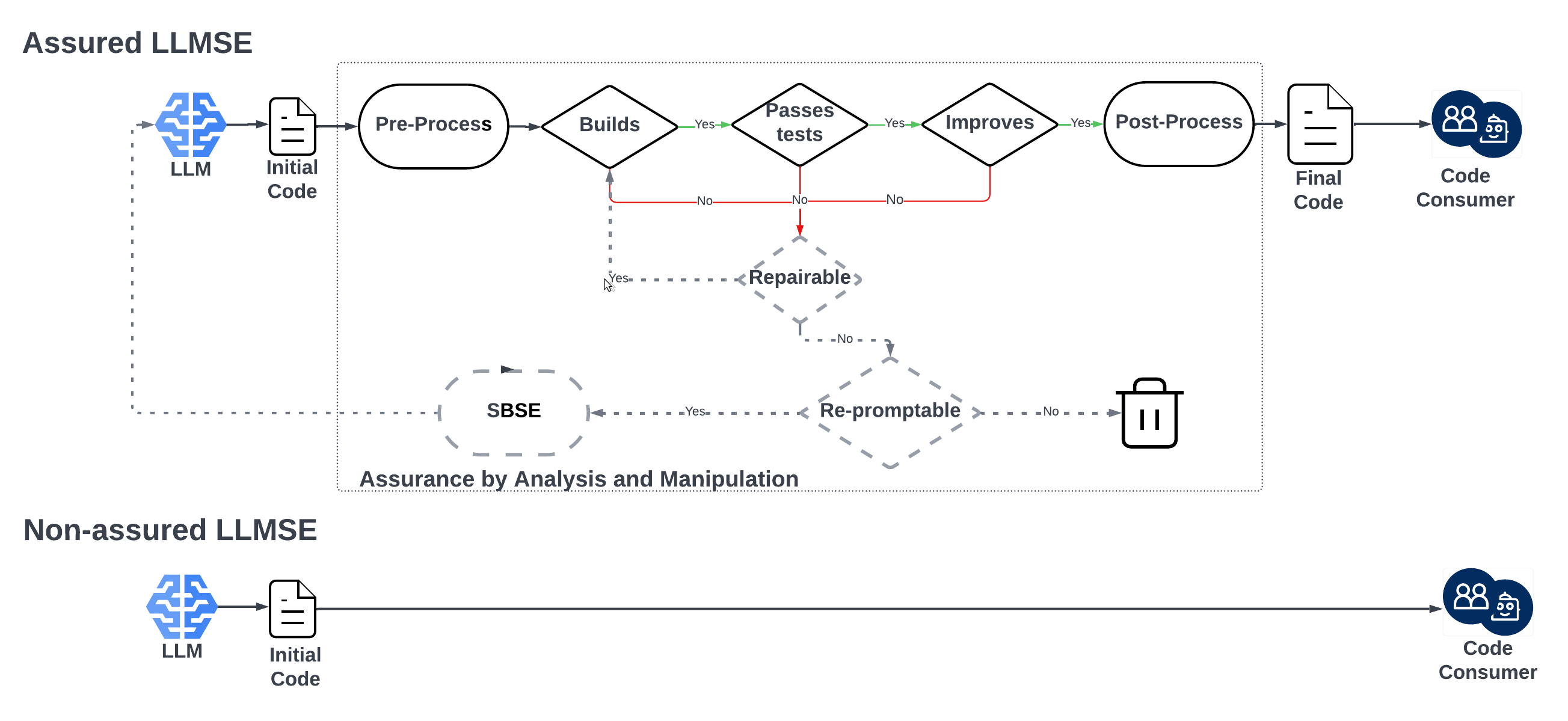}}
\vspace{-5mm}
\caption{Top level comparison between Assured and Non-Assured Large Language Model Software Engineering. {\rm In the assured mode, there is a whole infrastructure phase for implementing `Assurance by Analysis and Manipulation'. 
This  assurance phase pre-processes and post-processes the initial code produced by the language model, passing it through a set of filters. 
Code that passes through all these filters meets the measurable  assurance guarantees denoted by the filters and is passed on to the code consumer, which may be a human or another automated tool. 
Code that fails any of the filters may undergo a follow-up repair process, and/or may occasion LLM re-prompting and prompt optimisation using SBSE. 
The repair and re-prompting steps are optional. 
In general, code that fails any of the filters (and cannot be repaired or re-prompted) will be discarded. 
By comparison, Non-Assured LLMSE simply passes the initial code generated in response to an LLM prompt directly to the code consumer, and offers no guarantee; the code may not even compile.}
\label{fig:top_level}}
\end{figure*}

Figure~\ref{fig:top_level}  contrasts the assured approach (in the upper part of the figure) from the non-assured approach (in the lower part of the figure).
In the assured LLMSE approach the code initially produced by the LLM goes through pre-processing and post-processing and passes through a number of filters.
The filters depicted in Figure~\ref{fig:top_level} are merely illustrative. 
Any syntactic or semantic filter could be used, subject to the specific Assured LLMSE application of interest.
The filtration process gives the assurance that the final code that reaches any code consumer meets the requirements of the assurances.

Assured LLMSE can be either offline or online.
However, the provision of the required assurances may require considerable computational resources, including elapsed time, to compute them.
Therefore, many assured LLMSE applications may inherently reside only in the offline space.

For example, suppose code  is constructed with respect to a specification, and the code produced is formally proved correct.
This is a very stringent form of assured LLMSE, in which the code produced is fully verified.
Such formal verification typically comes at high computational cost \cite{hasan:formal}, making such a form of assured LLMSE likely unachievable online.
However, recent advances in LLM-based verification~\cite{first:baldur} may reduce this cost.
By comparison with formal verification, a weaker, but more readily-achievable form of assurance can be provided by  software testing.

\section{Assured Offline LLM-Based Software Engineering  as Genetic Improvement }
According to the Genetic Improvement (GI) \cite{Petke:gisurvey} approach, software is treated as `genetic material' to be `mashed up and recomposed' using source code analysis and manipulation \cite{mh:scam10}, and/or genetic programming \cite{koza:book} as an operator to generate candidate solutions.
GI  is a subclass of Search Based Software Engineering (SBSE) \cite{mhbj:manifesto,mhamyz:acm-surveys} in which the search space is the space of programs, and the fitness function that guides the search measures whether the candidate solutions improve on some set of characteristics of the original program.
GI has already proved widely applicable with optimization goals and applications
including 
execution time \cite{blmh:tec1,white:evolutionary-improvement}, 
power consumption \cite{bbetal:approximate,white:searching,monotas:seeds},
graphical rendering \cite{sitthi:genetic},
dynamic memory consumption \cite{fan:gecco15}, 
and
online safety \cite{ahlgren-etal:wes}.

The approach is often characterized as generate-and-test, especially when applied to program repair \cite{legoues:cacm-survey}.
A great many candidate solutions are generated.
Each is tested to check whether it meets the criteria for improvement.
Assured LLMSE can be thought of as a kind of Genetic Improvement in which LLMs are used as the operator for generating candidate solutions.
The filters we describe in this paper can  be thought of as playing a similar role to fitness functions for generate-and-test approaches to GI.

\subsection{SBSE Over the Prompt Search Space}
Filters tend to be boolean; either the candidate code passes through the filter or it is discarded.
However, for code that does not pass the filter, we can define the {\em degree} to which it fails thereby generalising the (boolean returning) filter to a (real-number-returning) fitness measurement.
Thinking of filters as fitness functions, and Assured LLMSE as GI, the backedge from the decision to re-prompt to the LLM in  Figure~\ref{fig:top_level} becomes most interesting.
It denotes the feedback from fitness computation to construction of suitable next prompt to the LLM(s).

Suppose the language model uses a chain of thought \cite{wei:chain}, which has already proved attractive for LLMSE \cite{mhetal:LLM-survey}.
This chain of prompts is the input to the LLM.
It captures the desired optimization goal.
Suppose further that, as well as sequencing of prompts in this chain, we also allow selection and repetition operators.
This kind of chain is thus defined using a Turing-complete programming language, in which programs express improvement goals.
The `re-prompt' backedge now becomes a move in the search space defined by this language.

We can now use techniques associated with traditional SBSE on simple programs in a language that expresses optimization goals in a format amenable to an LLM.
In this way, we can automatically optimize  prompts using computational search, for which the fitness is measured as the ability to pass the filters.
As a result, the overall Assured LLMSE process produces, not only code backed by verifiable assurances but, as an additional  byproduct, it continually optimizes its own prompting strategy.
This evolving prompting strategy is expressed in a simple domain-specific prompting language making it both human-readable and machine-optimizable.

\subsection{Assured Offline LLM-Based Performance Optimisation}
The greatest impact achievable from the Assured LLMSE agenda is likely to come from performance optimization. 
There are two distinct cases, which we characterize as “local” and “global”. 
Local performance optimization is a good starting point, because it will yield insight and impact. 
However, we anticipate greater impact (but also greater technical challenges) with global optimization.

\subsubsection{Local performance optimization} 
In local performance optimization, we limit the context required by the LLM, thereby increasing its chance to find optimizations, but reducing the scope, and therefore the likely impact on performance of these optimizations. 
Interviews with stakeholders at Meta suggested that local optimization opportunities may already largely be fully exploited; the company has many code modifications (codemod) services and rule-based systems that seek to apply local optimisation rules to tighten up on code performance.
Other code bases may be locally less well optimized than those at large tech companies such as Meta. 
This also makes local performance optimisation and natural target for assured offline LLMSE.

\subsubsection{Global performance optimization}
With global optimization, we would provide the full stack context to the language model, 
thereby allowing it to make several interrelated changes at once. 
These changes might modify different languages, systems and parts of the overall code stack.

Of course such global optimization may require a  large LLM token window size.
Global optimization will also likely prove challenging for an automated approach using LLMs, because this is the so-called “multi hunk problem” , which has so far proved challenging in closely related areas like automated program repair \cite{saha:harnessing}. 

Global optimization will also likely require sophisticated prompt formulation, using Chain of Thought (CoT) \cite{wei:chain} and Re-Act-style deployment \cite{yao:react}. 
CoT can be used to break the overall optimization problem down into more manageable steps, guiding the language model with chains of reasoning.

The ReAct approach can further inform the overall process with the ability to call out to ground truth.
For example, a CoT step might seek to explore the execution time impact of particular code change, with the ReAct approach providing a call to a execution environment API, that provides the ground truth on this execution time.
Using SBSE (see Figure~\ref{fig:top_level}), the prompt can, itself,  be thought of as a program (in a domain-specific prompting language) that is optimized by computational search.
In this way, as the overall workflow executes on different optimization tasks, it produces an optimized prompt strategy as a byproduct of its use.

Global performance optimization cannot be found with traditional 
automated code modification techniques (confined to specific performance patterns). 
It is also especially challenging for {\em engineers} to find optimizations that involve multiple `moving parts' in this way, 
because of the inherent  complexity and the requirement for knowledge of multiple domains and systems.
Nevertheless, because such optimizations are more far-reaching they can profoundly impact performance, making this a worthy topic for global LLMSE.

\subsubsection{Refactoring}
Surprisingly, refactoring is relatively under explored in the field of LLMSE \cite{mhetal:LLM-survey}.
This is surprising, because refactoring would prove to be a relatively less challenging target for LLMSE.
The assurance is relatively straightforward because the refactored code's functional behavior is expected to mimic that of the original.
Therefore, the assurance criteria can be fully automated using regression tests as an automated test oracle \cite{mhetal:LLM-survey,alshahwan:software}.
It is also less challenging because it can be effective in offline mode; large scale refactoring could be implemented over several hours or even days and yet still be exceptionally useful to the consumer, because it might otherwise take weeks or even months for human-led engineering effort to achieve the same result.

There are two ways in which LLMs can automate refactoring:  through known industry-standard design patterns \cite{gamma:design}, and through bespoke (project-specific) refactoring scale outs.

\medskip
\noindent
{\bf Automated transformation to incorporate best practice design patterns:}
Since they are trained on large code corpora, LLMs inherently have the potential to implement migration of existing code to implement a new design pattern, such as MVVM \cite{anderson:model}.  
However, simply applying the technology out of the box, at scale, could lead to a large amount of buggy code, due to hallucination.

We therefore need automated testing results as a ground truth correctness oracle in the Assured LLMSE filtration process. 
Testing needs to have high coverage to provide a strong oracle.
Testing needs to be adaptive, so that we can generate new tests for the LLM’s newly generated code.
Fortunately, automated test generation is a widely studied area and its own right with many candidate solutions \cite{anand2012automated,mh:icst15-keynote}.
Once an organisation has deployed automated testing to sufficiently high levels of test effectiveness, 
this will unblock the more widespread application of Assured  LLMSE for Refactoring.  

The beauty of the refactoring use case is that, by definition, refactoring is {\em expected} to respect all test cases designed to capture functional regressions (the Automated Regression Oracle \cite{mhetal:LLM-survey}). 
Therefore, Assured LLMSE  for refactoring dovetails well with automated regression testing. 
The LLM plays the role of codemod candidate generation, while the regression tests play the role of filter. 
Because tests are generated, not human-written, they can be generated to cover the newly generated code from the LLM. 

\medskip
\noindent
{\bf Bespoke refactoring:}
Engineers often need specific project-defined refactorings, such as changes to the way they call a particular API.
It is typically painful and slow to implement changes across all code in a large project \cite{mketal:evaluating-repair}.  
This can lead to software teams necessarily choosing to build up high levels of technical debt \cite{matheus:ase17}, 
rather than “biting the bullet” and performing the refactoring.  

Fortunately, LLMs have proved very effective at so-called “few shot learning” for Software Engineering problems \cite{mhetal:LLM-survey}. 
The implication of few shot learning for automated bespoke refactoring, is that  engineers would  need to provide just one or two example cases (such as changes to an API interface). 
From these examples, the LLM can be expected to automate the process of applying this re-factoring style across the project.  
Once again, the Automated Regression Oracle will be key as a source of test assurance, without which we cannot rely on the raw output of the LLM.

\subsubsection{Debugging} 
Debugging is expensive and tedious. 
If we can free engineers from this drudgery, 
they will not only be delighted, but free to work on more exciting development work. 
In the past year, the scientific literature on LLM-based automated repair (aka bug fixing) has developed rapidly \cite{mhetal:LLM-survey}. 

Through Chain of Thought \cite{wei:chain}, SBSE prompt program optimisation, and Re-Act-style deployment \cite{yao:react}, 
it may be possible to automate the process of root causing bugs. 
Automated bug fixing techniques have been widely studied in scientific literature, with good results reported, but they rely on high quality test coverage \cite{legoues:cacm-survey}.
However, despite initial industrial deployment \cite{ametal:sapfix}, there remain challenges to more widespread industrial uptake \cite{mh:apr22-keynote}.

\section{Open Research Problems}
This  position paper has set out the case for Assured Large Language Model-based Software Engineering (Assured LLMSE).
The provision of the assurances may require an offline deployment, since stringent assurances cannot be computed in real time. 
Nevertheless, the automatic discharging of all such assurance responsibilities makes Offline Assured LLMSE an attractive overall deployment model for LLM-based code generation.
Where the assurances can be computed in real time, so much the better.

We conclude by outlining some open problems for this research agenda, which brings together SBSE, GI and LLMSE.

\medskip
\noindent
{\bf Migrating Offline to Online LLMSE}:
Clearly online LLMSE would be preferable to offline but in the Assured LLMSE paradigm, it may not be possible to always migrate from offline to online. 
The offline paradigm also provides a suitable starting place for research. 
What is offline this year can perhaps be migrated to online next year.
In the meantime, the more relaxed constraints of offline mode may make it easier to focus on the development novel approaches, without having to optimise early for timing constraints.

More work is also needed to understand how best to migrate as many of the assurances provided with Assured LLMSE to the real time (online) deployment mode.

\medskip
\noindent
{\bf Defining effective and efficient filters for Assured LLMSE}:
The filters play the role of fitness functions, and therefore the entire field of software measurement becomes relevant and applicable \cite{maja:metrics}.
However, as has been noted in the field of SBSE \cite{mh:wetsom-keynote},
some traditional software measurement techniques may be too computationally expensive for use as fitness functions;
we need fast approximate metrics that can act as sufficiently accurate proxies.
We can draw an inspiration here from the field of Genetic Improvement.
For example, we can use increasingly large subsets of the available test suite as the search process matures from exploration to exploitation \cite{legoues:cacm-survey}.

\medskip
\noindent
{\bf Defining a sufficiently rich prompting language, and genetic operators over it}:
There is a large, as yet unexplored, green field research site located over the set of problems concerned  with the definition of prompting languages and the genetic operators on these languages.
The prompting language  needs to be sufficiently rich that is can capture effective LLMSE optimization strategies.
It also needs to be amenable to computational search, such as genetic programming. 
For example, we need operators for manipulating programs in the language that yield search landscapes in which computational search techniques can optimize.
There are considerable grounds optimism here, since what is needed is a very domain-specific language, and therefore it can be designed with all of these objectives in mind.
Traditional programming languages have tended to be less amenable to computational search, in part because they have needed to be highly general purpose.

\medskip
\noindent
{\bf Finding scalable ways to hybridize computational search and LLMSE}:
Locating SBSE at the heart of the optimization of prompting language programs makes sense conceptually: theoretically, it offers tremendous opportunities to optimize  prompting strategies.
Nevertheless, computational cost will pose a  barrier to practical deployment, even with fast approximate filter/fitness functions.
Therefore, we also need to find scalable hybridization approaches, that exploit the `embarrassing parallelism' \cite{herlihy:art} of both LLM response processing, and prompting language optimization.
In a fully parallel approach, multiple different responses from multiple different prompt programs will be evaluated simultaneously, thereby allowing the entire approach to scale.

\medskip
\noindent
{\bf Finding domain-aware code-aware computational search strategies}:
Much of the existing work on LLMSE has used off-the-peg language models in their default settings \cite{mhetal:LLM-survey}.
This neglects the huge potential to better exploit the LLM's conditional probability distribution.
Standard, one-size-fits-all approaches that collapse the distribution into a single response are inherently sub optimal.
We need to define strategies to navigate the probability distribution that are not only code-aware, but also domain-aware.
That is, a strategy will perform optimally well for a particular software engineering problem domain if it is defined with respect to the characteristics of that domain and also the nuances of the programming language in which responses are expressed.

\newpage

\balance

%
\bibliographystyle{ACM-Reference-Format}
\bibliography{main}

\end{document}